\documentstyle[prl,aps,floats,psfig]{revtex}
\begin{document}
\draft

\twocolumn[
\hsize\textwidth\columnwidth\hsize\csname@twocolumnfalse\endcsname

\title{Resistively-shunted superconducting quantum point contacts}  

\author{D.V. Averin, A. Bardas, and H.T. Imam} 

\address{Department of Physics and Astronomy, SUNY at Stony Brook, Stony Brook 
NY 11794}

\date{\today}
\maketitle

\begin{abstract}
We have studied the Josephson dynamics of resistively-shunted 
ballistic superconducting quantum point contacts at finite 
temperatures and arbitrary number of conducting modes. Compared 
to the classical Josephson dynamics of tunnel junctions, 
dynamics of quantum point contacts exhibits several new features 
associated with temporal fluctuations of the Josephson potential 
caused by fluctuations in the occupation of the current-carrying 
Andreev levels. 

\end{abstract}
]

Andreev levels with energies below the superconducting energy gap 
$\Delta$ are known to carry the stationary supercurrent in 
ballistic superconducting point contacts \cite{ko1,fur,been}, and 
are also responsible for the dc and ac current flow at small bias 
voltages $V\ll \Delta/e$ \cite{we1,we2,sh}. Occupation factors 
of these states vary randomly, either due to the thermal 
fluctuations in the contact electrodes \cite{n1,mr}, or due to 
the probabilistic nature of the Landau-Zener transitions 
\cite{n2} induced by the bias voltage in contacts with finite 
reflection coefficients \cite{we1}. These random fluctuations 
in the occupation factors of the sub-gap states lead to a large 
supercurrent noise. At finite bias voltages, the noise can be 
interpreted as the shot noise of the large charge quanta of 
magnitude $2\Delta/V$ which is equal to the charge transferred 
through the contact during one period of the Josephson 
oscillations. Recently, this noise has been observed in 
experiments with high-transparency tunnel junctions \cite{nexp}.   

In a point contact with an external environment of finite 
impedance, the supercurrent noise leads to fluctuations of the 
voltage across the contact, and affects the dynamics of the 
Josephson 
phase difference $\varphi$. In the case of quantum point contacts 
with few propagating electron modes, for example those fabricated 
with the controllable break junction technique \cite{van,oub,sac}, 
the typical external impedance is much smaller than the contact 
resistance. In this case the effects of the external impedance 
are significant only at small bias voltages, where dynamics of 
$\varphi$ can be described within the adiabatic approach 
\cite{we2}. The aim of this work is to study adiabatic 
finite-temperature dynamics of superconducting quantum point 
contacts, taking into account the supercurrent noise caused by 
fluctuations in the occupation factors of the Andreev levels. 

We consider the standard model of a point contact, characterized 
as a short ballistic constriction supporting $N$ electron modes 
propagating between two identical bulk superconductors. At 
small bias voltages, $V\ll \Delta/e$, the current through such 
a constriction is carried by the two levels (per each electron 
mode) with energies $\pm \epsilon (\varphi)$, where 
$\epsilon(\varphi) = \Delta \cos (\varphi/2)$. 
We assume that the external impedance of the contact 
reduces to the frequency-independent constant $R$ in the relevant 
frequency range around the typical Josephson oscillation frequency 
of $2eI_cR/\hbar$, where $I_c= N\Delta e/\hbar$ denotes the 
zero-temperature critical current of the contact. 

The basic result of the adiabatic theory \cite{we2} of dynamics 
of superconducting point contacts is that each occupied Andreev 
state contributes energy $\epsilon(\varphi)$ or $-\epsilon(\varphi)$ 
to the Josephson coupling energy of the point contact, while 
the occupation probability $w$ of the state is determined by 
the quasiparticle transitions between this state and the 
electrodes of the contact. These transitions are described by 
the simple rate equation:  
\begin{equation} 
\dot{w}=  \gamma(\epsilon) [f(\epsilon)-w] \, ,
\label{0} \end{equation} 
where $f(\epsilon)$ is the equilibrium Fermi distribution, 
$\epsilon=\epsilon(\varphi(t))$, and the transition rate 
$\gamma(\epsilon)$ depends on the inelastic or pairbreaking 
effects which give rise to the finite subgap density of states 
in the electrodes \cite{we2}. Equation (\ref{0}) shows that the  
role of quasiparticle transitions is to drive the occupation 
probabilities of Andreev states towards the equilibrium.  

The time evolution of the Josephson phase difference $\varphi(t)$ 
affects the occupation probabilities of Andreev states through 
the phase dependence of their energies $\pm \epsilon(\varphi)$. 
At finite temperatures the phase dynamics is diffusion along the 
Josephson potential driven by the thermal noise generated by the 
external resistance $R$. Such a diffusion is governed by the 
following set of Fokker-Planck equations for the probability 
density $\sigma$: 
\begin{eqnarray}
\dot{\sigma}_k (\varphi,t) +\frac{\partial j_k}{\partial 
\varphi} = \Gamma \{ \sigma_k \} \, , \;\;\;\;\nonumber\\ 
j_k (\varphi) = - (\frac{2e}{\hbar})^2 R \left[ T  
\frac{\partial \sigma_k }{\partial \varphi} + \sigma_k 
\frac{ \partial U^{(k)}}{\partial \varphi}  \right] \, , 
\label{1} \end{eqnarray}   
where $k=-N,-N+1,...,N$ is the difference of the numbers of 
the occupied Andreev levels that carry positive and negative 
currents. It characterizes different realizations of the 
Josephson potential, $U^{(k)} =-k\epsilon (\varphi)-(\hbar I_0/
2e)\varphi$, where $I_0$ is the external bias current (see Fig.\ 
1). The relaxation term $\Gamma \{ \sigma_k \}$ in eq.\ (\ref{1}) 
describes the transitions from one realization of Josephson 
potential to another, induced by the quasiparticle 
transitions (\ref{0}) between the Andreev states and the 
electrodes of the point contact. Taking a sum over transitions 
from, and into, all of $2N$ Andreev levels we get for 
$\Gamma$:  
\begin{eqnarray}
\Gamma \{ \sigma_k \} = \gamma(\epsilon) f(\epsilon) 
\left[ (N+k+1) \sigma_{k+1} -(N+k) \sigma_k \right] + 
\nonumber \\ 
\gamma(\epsilon)f(-\epsilon) \left[ 
(N-k+1) \sigma_{k-1}-(N-k) \sigma_k \right] \, . 
\label{2} \end{eqnarray}   

\begin{figure}
\centerline{
\psfig{file=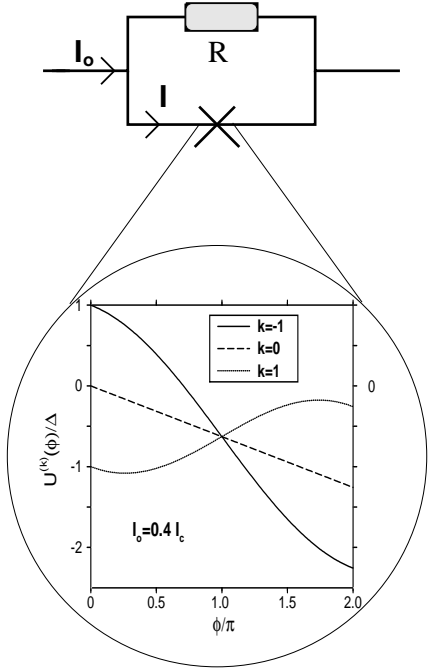,height=3.5in,angle=0}}
\caption{ 
Equivalent circuit of the resistively shunted 
superconducting quantum point contact, and three realizations 
of the Josephson potential for a single-mode ballistic 
contact. The phase diffusion along this potential is 
characterized by the three-component probability density 
$\sigma_k(\varphi)$ which should be matched continuously 
between the points $\varphi=0$ and $\varphi=2\pi$ \protect 
-- see eqs.\ (\ref{8}) and (\ref{9}). 
}\label{f1} 
\end{figure}

 Since the dependence of all ``internal'' properties of the 
point contact (for instance, Andreev state energies $\pm \epsilon 
(\varphi)$) on $\varphi$ is periodic with the period $2\pi$, it 
is convenient to restrict $\varphi$ to the interval $[0,\,2\pi]$. 
The diffusion equation (\ref{1}) should then be complemented 
with the boundary conditions for $\sigma$'s at the ends of this 
interval, $\varphi= 0, \,2\pi$. The boundary conditions are 
somewhat complicated by the fact that when the phase $\varphi$ 
approaches 0 or $2\pi$ the energies of the Andreev state hit 
the gap edges $\pm \Delta$. As a result, the quasiparticle 
transitions between these states and electrodes become 
``absolutely'' efficient, leading to instantaneous 
equilibration of the occupation probabilities of Andreev states 
\cite{we2}. (More accurately, one should say that this 
equilibration process has a time scale that is negligible in 
comparison to the period of the Josephson oscillations.) Since 
different Andreev states are occupied independently one from 
another, this implies for the probability densities:
\begin{equation}
\sigma_k (\varphi= 0) = A [f(-\Delta)]^{N+k} [f(\Delta)]^{N-k} 
C_{2N}^{N+k} \, ,   
\label{3} \end{equation}
\begin{equation} 
\sigma_k (\varphi= 2\pi)=A' [f(\Delta)]^{N+k} [f(-\Delta)]^{N-k} 
C_{2N}^{N+k} \, ,  
\label{3*} \end{equation}
where $A,A'$ are some constants, and $C_n^m=n!/[m!(n-m)!]$.  

Relations (\ref{3}) and (\ref{3*}) can serve as the boundary 
conditions for the diffusion equations (\ref{1}). One more 
condition follows from the requirement that the components 
$\sigma_k (\varphi)$ of the probability density diffusing on 
the different branches $U^{(k)} (\varphi)$ of the Josephson 
potential be continuous. Taking into account that the states of 
the point contact with $\varphi$ and $\varphi+2\pi$ should be 
identical, we can write this condition as $A'=A$. Combining 
this result with eqs.\ (\ref{3}) and (\ref{3*}), we see that 
one of this equations can be replaced with the following
boundary condition:  
\begin{equation} 
\sigma_k (\varphi=2\pi) = \sigma_{-k} (\varphi=0) \, . 
\label{5} \end{equation}

Equations (\ref{3}) and (\ref{5}), together with the usual 
normalization condition  
\begin{equation}
\sum_k \int_0^{2\pi} d \varphi \sigma_k (\varphi) =1 \, ,
\label{4} \end{equation}
determine completely the solution of the diffusion equation 
(\ref{1}), and thus allow us to find all the characteristics 
of the point contact. Particularly, in the steady-state case, 
eq.\ (\ref{1}) 
reduces to the equation $\partial j_k/\partial \varphi = 
\Gamma \{ \sigma_k \}$, and can be solved explicitly to 
obtain the dc voltage $V$ across the point contact: 
\begin{equation}
V=\frac{\pi\hbar}{e} \sum_k j_k \, . 
\label{6} \end{equation}
  
Equations (\ref{1}) -- (\ref{5}) show that the time evolution of 
the $\varphi$ depends, in a non-trivial way, on the number $N$ 
of the propagating electron modes in the point contact. Below we 
analyze two opposite limits of such a dependence, $N=1$ and 
$N\gg 1$. We also assume that the subgap density of states in 
the electrodes of the point contact is small and we can neglect 
the quasiparticle exchange term $\Gamma$ in eq.\ (\ref{1}). We 
start with the case of a {\em single-mode quantum point contact}  
with $N=1$. We have then three diffusion equations (\ref{1}) 
with $k=0; \, \pm 1$ (see Fig.\ 1). The boundary conditions 
(\ref{3}) and (\ref{5}) are: 
\begin{equation}
\sigma_1(0)= \frac{1}{2} \sigma_0(0) e^{\Delta/T}
\, , \;\;\;  \sigma_{-1}(0)= 
\frac{1}{2}\sigma_0(0) e^{-\Delta/T} \, , 
\label{8} \end{equation}   
and 
\begin{equation}
\sigma_{\pm 1}(2\pi)=\sigma_{\mp 1}(0) \, , \;\;\; 
\sigma_0 (2\pi)=\sigma_0 (0) \, .
\label{9} \end{equation}   

The stationary version of eq.\ (\ref{1}), $j_k(\varphi)=
\mbox{const}$, can be integrated directly, as in the case 
of classical Josephson junctions \cite{iz,ah}. Taking into 
account the boundary conditions (\ref{8}) and (\ref{9}) we 
find then the $\sigma$'s: 
\begin{eqnarray} 
\sigma_{\pm1} (\varphi)= &\frac{\sigma_0}{2}& \exp \{-\frac{U^{(\pm1)}  
(\varphi)}{T} \} [ 1-\frac{1}{K_{\pm}} (1-e^{-\pi \hbar I_0/eT}) 
\nonumber\\
&\times& \int_0^{\varphi}d\varphi'  \exp \{\frac{U^{(\pm1)} 
(\varphi')}{T} \} ]\, ,
\label{10} \end{eqnarray}
where $K_{\pm}=\int_0^{2\pi}d\varphi  \exp \{U^{(\pm1)}(\varphi)/T\}$, 
and $\sigma_0$ is a constant density $\sigma_0(\varphi)$ that 
is determined by the normalization condition 
\begin{equation} 
2\pi \sigma_0 +\sum_{\pm1} \int_0^{2\pi}d\varphi \sigma_{\pm1} 
(\varphi)= 1 \, . 
\label{11} \end{equation} 
In terms of $\sigma_0$, the dc voltage $V$ across the point 
contact is: 
\begin{equation}
V=2 \pi \sigma_0 I_0 R \left[ 1+\frac{eT}{\hbar I_0}(\frac{1} 
{K_+} +\frac{1}{K_-})(1-e^{-\pi \hbar I_0/eT})\right] \, . 
\label{12} \end{equation} 

Figure 2 shows the current $I$ flowing through the point 
contact, $I\!=\!I_0\! -\!V/R$ (relation between $I$ and 
$I_0$ is illustrated by the equivalent circuit in Fig.\ 1), 
as a function of the voltage $V$ calculated from eqs.\ 
(\ref{10})--(\ref{12}) for several temperatures. The 
zero-temperature $I-V$ characteristic (the upper curve in Fig.\ 
2) can be represented analytically by the following dependence 
of the voltage $V$ on the bias current $I_0$ \cite{we2}: 
\begin{equation} 
V=\frac{\pi R(I_0^2 -I_c^2)^{1/2}}{4\arctan \sqrt{ 
(I_0+I_c)/(I_0-I_c)} } \, , \;\;\;\; I_0>I_c \, , 
\label{13} \end{equation}
where $I_c=e\Delta/\hbar$ is the zero-temperature supercurrent 
of the point contact. This equation describes the transition from 
the supercurrent at $V=0$ to a constant current $2I_c/\pi$ at 
$V\gg RI_c$. The large-voltage dc current arises from the 
strongly non-equilibrium occupation of the Andreev levels,  
and is a hallmark of multiple Andreev reflections in a ballistic 
point contact. 

Both the zero-voltage supercurrent and the large-voltage current 
are suppressed by temperature. For instance, in the limit of 
large temperatures, $T\gg \Delta$, we get from eqs.\ 
(\ref{10})--(\ref{12}): 
\begin{equation} 
I=I_0 \left( \frac{\Delta}{2T}\right)^2 \left[\frac{1}{1+i^2} 
+\frac{4i^3}{\pi(1+i^2)^2} \coth (\frac{\pi i}{2}) \right] \, 
i\equiv \frac{\hbar I_0}{eT} \, . 
\label{14} \end{equation}     
This equation describes well the lowest curve in Fig.\ 2. 
Transition from low to high temperatures can be traced 
analytically at large bias currents, $I_0\gg I_c$, when 
eqs.\ (\ref{10})--(\ref{12}) give:
\begin{equation}
I=\frac{2}{\pi} I_c \tanh (\frac{\Delta}{2T})\, , 
\label{15} \end{equation} 
in agreement with Fig.\ 1 and large-current limit of eq.\ 
(\ref{14}). 
 
At small bias currents $I_0<I_c$, and small temperatures $T\ll 
\Delta$, the point contact is in the supercurrent state and the 
bias current flows almost completely through the contact, 
$I\simeq I_0$. The voltage across the contact is exponentially 
suppressed and is associated with the rare ``phase-slip'' 
events $\varphi \rightarrow \varphi +2\pi$. In this regime, and 
for bias currents on the order of the critical current $I_c$, we 
get from eqs.\ (\ref{10})--(\ref{12}): 
\begin{eqnarray}
V=\frac{RI_c}{2} (1- i^2)^{1/2} \exp\{- \frac{2\Delta}{T} 
[ (1- i^2)^{1/2} \nonumber\\
-i(\frac{\pi}{2} -\sin^{-1}(i)) ] 
\} \, , \;\;\;\; i\equiv \frac{I}{I_c}. 
\label{16} \end{eqnarray}
This equation corresponds to the classical thermal activation 
over the maximum of the Josephson potential, similar to the one 
found in classical Josephson junctions \cite{iz,ah}. 

\begin{figure}
\centerline{
\psfig{file=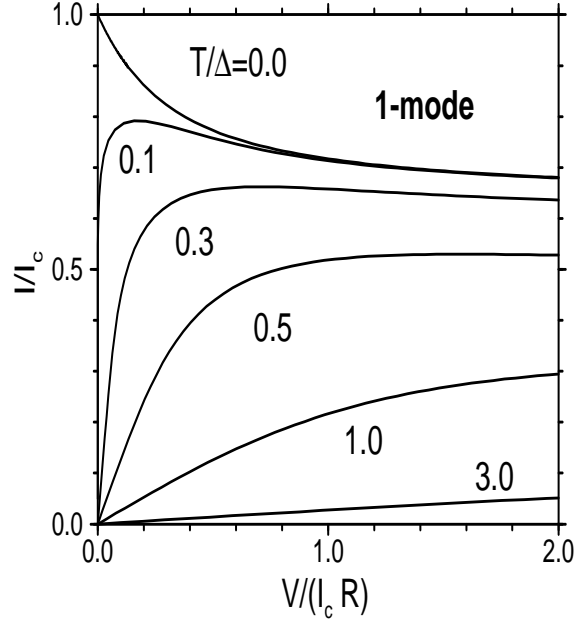,height=3.5in,width=3.2in,angle=0}}
\caption{
DC $I-V$ characteristics of the single-mode 
superconducting ballistic quantum point contact at several 
temperatures. All components of the current are suppressed 
when temperature $T$ increases above the energy gap 
$\Delta$. The various limits of these curves are given 
analytically \protect by eqs.\ (\ref{13}) -- (\ref{16}). 
}\label{f2} 
\end{figure}

Now we consider {\em point contacts with large number of 
transverse modes}, $N\gg1$. The probability density $\sigma_k$ 
is strongly peaked in this case as a function of $k$ at its 
average value: $k=(2w-1)N$, where $w$ is the occupation 
probability of the Andreev states carrying positive current. 
This means that we can neglect fluctuations of the Josephson 
potential around the average potential:
\begin{equation}
U(\varphi)= N\epsilon (\varphi)(1-2w) \, .
\label{17} \end{equation} 
The probability $w$, which determines the shape of the potential 
(\ref{17}), depends sensitively on the equilibration rate 
$\gamma$ in the rate equation (\ref{0}). We start by considering 
the case of an ideal BCS superconductor in which $\gamma$ is 
completely suppressed in the relevant sub-gap energy range, and 
equilibration of the Andreev states occurs only at 
$\varphi=0,\, 2\pi$. In this case the probability $w$ is equal 
to $f(-\Delta)$ or $f(\Delta)$ depending on whether the phase 
diffuses into the interval $[0,\,2\pi]$ through one ($\varphi 
=0$), or the other ($\varphi=2\pi$), end of this interval.  
This means that although we neglect fluctuations of the 
Josephson potential, we still have two branches of this 
potential due to the non-equilibrium occupation of Andreev 
states. The potential is $E_J \cos (\varphi/2)$ on one branch, 
and $-E_J \cos (\varphi/2)$ on the other, where $E_J\equiv 
N\Delta \tanh (\Delta/2T)$. The phase diffusion along these 
two branches can be described with the two equations in 
(\ref{1}) with $k=\pm 1$, if we replace the energy 
gap $\Delta$ in $U^{(k)}(\varphi)$ with $E_J$. However, the 
boundary conditions for $\sigma_{\pm 1}$ are now quite 
different from those we used for the single-mode junction 
-- see eqs.\ (\ref{8}) and (\ref{9}). As we discussed above, 
the phase diffuses along one or the other branch of the 
potential depending on whether it enters the interval 
$[0,\,2\pi]$ through one or the other end. This dynamics is 
accounted for by the following boundary conditions: 
\begin{equation}
\sigma_1(2\pi)=\sigma_{-1}(0) =0 \, , \;\;\; \sigma_{-1} 
(2\pi)= \sigma_1 (0) \, .
\label{17*} \end{equation}   

Equations (\ref{1}) with the boundary conditions (\ref{17*}) 
give: 
\begin{equation} 
\sigma_{1} (\varphi)= \frac{\sigma}{K_+} \exp 
\{-\frac{U^{(1)}(\varphi)}{T} \} \int_{\varphi}^{2\pi} 
d\varphi' \exp \{\frac{U^{(1)} (\varphi')}{T} \} \, ,
\label{18} \end{equation}
\vspace{-0.5in} 
\begin{eqnarray} 
\sigma_{-1} (\varphi)= \frac{\sigma }{K_-}e^{-\pi \hbar I_0/eT} 
\exp \{-\frac{U^{(-1)}(\varphi)}{T} \} \nonumber\\
\times \int_0^{\varphi}d\varphi' \exp \{\frac{U^{(-1)} (\varphi')}{T} 
\} \, ,
\label{19} \end{eqnarray} 
and 
\begin{equation}
V=\frac{4 \pi e R}{\hbar} \sigma T \left(\frac{1}{K_+} - 
\frac{e^{-\pi \hbar I_0/eT}}{K_-}\right) \, , 
\label{20} \end{equation} 
where $\sigma$ is a constant that is determined by the 
normalization condition (\ref{4}).

Figure 3 shows the $I-V$ characteristics of the multi-mode 
point contact calculated from eqs.\ (\ref{18})--(\ref{20}). 
The zero-temperature curve is given by the same eq.\ (\ref{13}) 
as for $N=1$, but with $I_c(T)=eE_J/\hbar=N(e\Delta/\hbar) \tanh 
(\Delta/2T)$. (Note that in the multi-mode contact, $E_J$ is 
much larger than $\Delta$, and the ``zero-tempertature'' 
$I-V$ characteristics with $T\ll E_J$ can still correspond to 
$T\gg \Delta$.) Since the current in Fig.\ 3 is normalized to 
the temperature-dependent critical current $I_c$, the 
large-voltage limit of the current in Fig.\ 3 is independent 
of temperature and equal to $2I_c/\pi$ in contrast to the 
single-mode case shown in Fig.\ 2. 

At large temperatures, $T\gg E_J$, the current $I$ can be found 
analytically from eqs.\ (\ref{18})--(\ref{20}): 
\begin{equation} 
I=\frac{2I_c}{\pi}\frac{i^2}{1+i^2}\coth (\frac{\pi i}{2}) 
\, , \;\;\; i\equiv \frac{\hbar I_0}{eT} \, . 
\label{21} \end{equation}     
This equation agrees with the lowest curve in Fig.\ 3. 
At low temperatures, $T\ll E_J$, and small bias currents 
$I_0<I_c$, the contact is in the supercurrent state and the 
voltage is generated only by the rare thermal activation 
over the Josephson energy barrier as in the single mode case. 
For not-too-small bias currents, the voltage is given by the 
same eq.\ (\ref{16}) with $\Delta$ replaced by $E_J$. At very 
small currents, $I_0/I_c\ll (T/E_J)^{1/2}$, the voltage is: 
\begin{equation}
V=2I_cR e^{-2E_J/T} \sinh (\frac{\pi \hbar I_0}{eT}) \, . 
\label{22} \end{equation} 

Finally, we qualitatively discuss the effect of the finite rate 
$\gamma$ of quasiparticle exchange between the Andreev states 
and contact electrodes. At small voltages $V\ll\hbar \gamma/e$ 
the phase evolves slowly and the quasiparticle transitions 
maintain equilibrium occupation of Andreev states. This means 
that the Josephson potential (\ref{17}) reduces to 
\begin{equation}
U(\varphi)=-N\epsilon (\varphi) \tanh (\frac{\epsilon (\varphi)
}{2T}) \, .
\label{23} \end{equation} 

\begin{figure}
\centerline{
\psfig{file=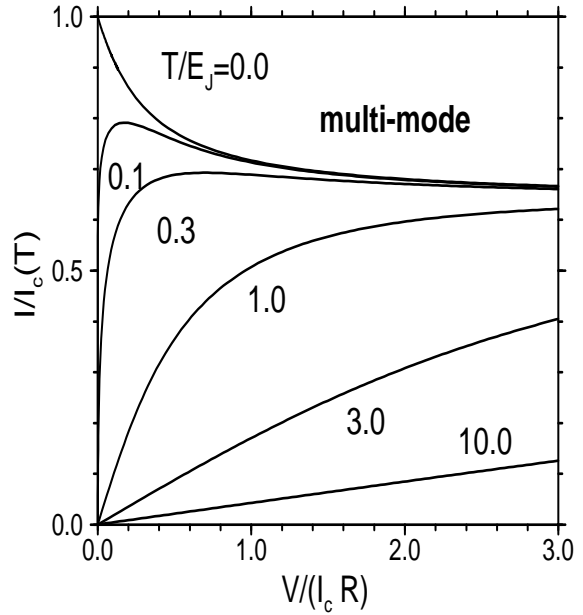,height=3.5in,width=3.2in,angle=0}}
\caption{ 
DC $I-V$ characteristics of the multi-mode quantum 
point contact between two BCS superconductors at several 
temperatures. For discussion see text. 
}\label{f3} 
\end{figure}

The main result of such a reduction is that now the potential 
depends only on the instantaneous value of $\varphi$, and 
therefore has only one branch instead of two as in absence of 
the quasiparticle relaxation. The exact shape of the potential 
(\ref{23}) depends on temperature, changing from $-N\Delta 
\mid \cos (\varphi/2)\mid$ at $T\ll \Delta$ to $-(N\Delta^2/4T)
\cos \varphi$ at $T\gg \Delta$. In this temperature range, 
$T\simeq \Delta$, the temperature is still negligible on the 
scale of the Josephson potential, and we can use the 
zero-temperature version of eqs.\ (\ref{1}) to describe the 
dynamics of $\varphi$. Combining this equation with the boundary 
condition  $\sigma(2\pi)=\sigma(0)$ appropriate for the 
situation with one branch of the potential, we get for the 
voltage $V$: 
\begin{equation}
V=2\pi R \left[ \int_0^{2\pi} \frac{d\varphi}{I_0 -I_s(\varphi)} 
\right]^{-1} \, , 
\label{24} \end{equation} 
where $I_s\equiv (2e/\hbar)dU(\varphi)/d\varphi$. In the two 
limits $T\gg \Delta$ and $T\ll \Delta$, eq.\ (\ref{24}) gives 
the standard result of the resistively-shunted-junction (RSJ) 
model, $V=R(I_0^2-I_c^2)^{1/2}$, where $I_c=$max$_{\varphi} 
[I_s(\varphi)]$. It can be shown by numerical integration of 
eq.\ (\ref{24}) that deviations from this result are smaller 
than a few percent for arbitrary $\Delta/T$ ratio. 

When the temperature becomes nonvanishing on the scale of 
Josephson potential, it is already much larger than $\Delta$ 
and the potential (\ref{23}) becomes the regular 
Josephson potential $-E_J \cos \varphi$. This means that 
the quasi-equilibrium dynamics of the multi-mode ballistic 
contacts can be described with the classical RSJ model 
\cite{iz,ah}. When the voltage $V$ across the contact 
is sufficiently large, $V\gg \hbar\gamma/e$, this 
quasi-equilibrium dynamics goes over into the non-equilibrium 
one, described by Fig.\ 3 and eqs.\ (\ref{13}), (\ref{21}).   

In summary, we have presented a new approach to the description 
of the classical Josephson dynamics of superconducting quantum 
point contacts with arbitrary number of conducting quantum modes 
in a resistive environment. New characteristic features of the 
Josephson dynamics in these contacts include fluctuations of 
the Josephson potential caused by fluctuations in the occupation 
of the current-carrying Andreev levels, and low-voltage excess 
current associated with the non-equilibrium occupation of these 
levels. 

We would like to thank A. Korotkov, K. Likharev, and S. Tolpygo 
for useful discussions. This work was supported in part by the 
ONR grant \# N00014-95-1-0762 and by the U.S. Civilian Research 
and Development Foundation under Award No.\ RP1-165.



%
%
%
%


\end{document}